\documentclass[11pt]{article}
\usepackage{amssymb,amsmath} 
\usepackage{graphicx}
\def\={\equiv} 
 
\def\pl{\partial}

\newcommand{\cast}{\circledast}

\newcommand{\bib}{\bibitem}

\newcommand{\nt}{\notag}
\newcommand{\ci}{\cite}
\newcommand{\lab}{\label}

\newcommand{\eq}{\eqref}

\newcommand{\bx}[1]{ \boxed{#1}}

\newcommand{\2}[1]{{\tilde #1}}

\newcommand{\9}[1]{^{\,\scriptscriptstyle#1}}

\newcommand{\w}{\hskip.5pt\9\flat}

\def\a{\alpha}

\def\e{\varepsilon} 

\def\f{\phi}

\def\m{\mu}

\def\r{\rho}
 
\def\s{{\sigma}}

\def\O{\Omega}

\def\S{\Sigma}

\newcommand{\hb}[1]{{\ \text{#1}\ }}
\newcommand{\bul}{$\bullet\ $}

\newcommand{\db}{{\,{\rm d}\kern-.9ex {^-}}\!}
\newcommand{\dir}{{\pl\kern-1.2ex {/}}}

\newcommand{\ie}{{\it ie, }}

\newcommand{\imp}{\ \Rightarrow\ }
\newcommand{\inv}{^{-1}}

\newcommand{\lra}{\leftrightarrow}

\newcommand{\plra}{\pl^{\kern-1.25ex^\lra}}
\newcommand{\qq}{\quad} 
\newcommand{\qqq}{\qquad}

\newcommand{\sr}{\sqrt}

\def\XXint#1#2#3{{\setbox0=\hbox{$#1{#2#3}{\int}$}
     \vcenter{\hbox{$#2#3$}}\kern-.5\wd0}}

\def\w{\wedge}

\def\bib#1{\bibitem[#1]{#1}}

\begin{document}

\title{Energy-momentum conservation in pre-metric electrodynamics with magnetic charges}

\author{Gerald Kaiser\thanks{
Research supported by AFOSR Grant \#F49620-01-1-0271.}\\
Center for Signals and Waves, Austin, TX\\
kaiser@wavelets.com  $\bullet$\ www.wavelets.com}

\maketitle

\begin{abstract}\noindent
A necessary and sufficient condition for energy-momentum conservation is proved within a topological,  pre-metric approach to classical electrodynamics including magnetic as well as electric charges. The extended Lorentz force, consisting of mutual actions by $F\sim(E, B)$ on the electric current and $G\sim(H, D)$ on the magnetic current, can be derived from an energy-momentum `potential' if and only if the constitutive relation $G=G\0F$ satisfies a certain vanishing condition. The electric-magnetic reciprocity introduced by Hehl and Obukhov is seen to define a complex structure on the tensor product $\2\O^{2,2}=\O^2\otimes\2\O^2$ of 2-form pairs $(F,G)$ which is independent of but consistent with the Hodge star operator defined by any Lorentzian metric. Contrary to a recent claim in the literature, it does \sl not \rm define a complex structure on the space of 2-forms itself.
\end{abstract}

\maketitle

\section{Pre-metric electrodynamics with magnetic charges}

In a recent monograph  \ci{HO3}, Hehl and Obukhov gave an axiomatic formulation of \sl pre-metric \rm classical electrodynamics. In that approach, the initial assumptions on spacetime are minimal: it is a differentiable 4-manifold $M$ foliated by a `topological time' parameter $\s$.  Maxwell's equations are expressed in terms of the  \sl field strength \rm 2-form $F\sim(E, B)$, the \sl field excitation \rm  2-form $G\sim(H, D)$, and the electric current 3-form $J\sim(j,\r)$ by
\begin{align}
dF&=0, \qq  F= B+E\w d\s\lab{F}\\
dG&=J, \qq  G= D-H\w d\s, \qq J=\r-j\w d\s,\lab{G}
\end{align}
where $(E, H)$ are 1-forms, $(B, D, j)$ are 2-forms and $\r$ is a 3-from, all spatial (pullbacks from the leaves).  $G$ and $J$ are \sl twisted \rm forms, variously called \sl densities \rm  \ci[page 183]{B85}, \sl pseudoforms \rm \ci{F1} or \sl odd \rm forms \ci{R84}, while $F$ is untwisted.  Additional spacetime structures, including light cones and other elements of Lorentzian spacetime, emerge only after further conditions are imposed via \sl constitutive relations \rm defining the medium in which the fields propagate.

The purpose of this note is to show that the magnetic flux conservation axiom in \ci{HO3}, represented above by the first equation \eq{F}, is unnecessary. To see this, note first that the only fundamental use made of $dF=0$ is in establishing the condition for energy-momentum conservation in \ci[Section B.5]{HO3}.  I will prove this condition \sl without \rm assuming $dF=0$ in the more general  setting of $n$ spacetime dimensions and fields represented by a $p$-form $F$ and an $(n-p)$-form $G$. We assume that $G$ and $J$ are twisted, though this will play no role in the proof. For $n=4$ and $p=2$, the conservation of magnetic flux is replaced by conservation of magnetic charge, whose current is represented by the 3-form $dF$.

Magnetic charges have been discussed extensively by Schwinger \ci{S75}, and conserved energy-momentum tensors have been constructed in the presence of magnetic currents by Rund  \ci{R77} and Moulin \ci{M1}. But to the best of my knowledge, the construction given here is the first that does not presuppose a spacetime metric or constitutive relations. I will establish a \sl necessary and sufficient condition \rm for electromagnetic energy-momentum conservation in the above setting. The choice of a Lorentzian metric then provides a sufficient, but not necessary, condition. 

Given a vector field $u$ on $M$, denote the contraction of $F$ by $u$ and its Lie derivation along $u$ by
\begin{align*}
u F\= u\rfloor F\,,\qq \5L_u F= d(uF)+u d F.
\end{align*}
Since the $(n+1)$-forms $F\w dG$ and $dF\w G$ must vanish, we have
\begin{align}\lab{sym}
u(F\w d G)&= uF\w d  G+\1p\, F\w u d  G=0, \hb{\ where\ } \1p\=(-1)^p\\
u( d  F\w G)&= u d  F\w G-\1p\,   d  F\w uG=0\nt.
\end{align}
Therefore
\begin{align}
\,d (F\w uG)&=\,d  F\w uG+\1p\,  F\w d(u G)\nt \\
&=\,d  F\w uG+\1p\,  F\w(\5L_u G-u\,d  G)\nt \\
&=\1p\,  F\w\5L_u G+\,d  F\w uG+uF\w \,d  G\nt \\
&=\1p\,  F\w\5L_u G+f_u \lab{a}
\end{align}
where
\begin{align}\lab{lorentz}
f_u\=\,d  F\w uG+uF\w \,d  G.
\end{align}
Similarly,
\begin{align}
\1p\,d (uF\w G)&= \1p\, (d(uF)\w G-\1p\,  uF\w \,d  G)\nt \\
&= \1p\, (\5L_u F-u\,d  F)\w G-uF\w \,d  G\nt \\
&=\1p\, \5L_u F\w G- d F\w uG- uF\w \,d  G \nt \\
&=\1p\, \5L_u F\w G-f_u\,.  \lab{b}
\end{align}
Taking the sum of  \eq{a} and \eq{b} gives
\begin{align*}
\,d (uF\w G+\1p\, F\w uG)=du(F\w G)=\5L_u F\w G+F\w\5L_u G=\5L_u(F\w G),
\end{align*}
which is trivial by the definition of $\5L_u$ since $u\,d (F\w G)=0$ as $F\w G$ is a  4-form.  Therefore the entire content of \eq{a} and \eq{b} is in the \sl difference, \rm which we write as
\begin{align}\lab{en-mom}
\bx{\ d\S_u=f_u+\f_u \ }
\end{align}
where
\begin{align}
\S_u&=\frac12(F\w uG-\1p\,  uF\w G)\lab{Su}\\
\f_u&=\frac{\1p}2(F\w\5L_u G-\5L_u F\w G).\lab{fu}
\end{align}
Specializing to $n=4$ and $p=2$, $F$ is interpreted as the field strength and $G$ as the excitation generated by the \sl electric and magnetic currents \rm
\begin{align}\lab{Max}
J\=d  G,\qqq K\=d  F.
\end{align}
The 3-form $\S_u$ is the local \sl kinematical energy-momentum field density \rm  and the 4-form $f_u$ is the  \sl extended Lorentz force density, \rm as defined in \ci{HO3} for the special case $K=0$.\footnote{$\S_u$ is `kinematical' because in the pre-metric setting there is insufficient structure to define a \sl dynamics. \rm This is somewhat analogous to a physical system described in \sl phase space \rm (a symplectic manifold) before a specific Hamiltonian is selected.
}
Equation \eq{en-mom} shows that $\S_u$ acts as a \sl potential \rm for $f_u$ if and only if $\f_u=0$.  That is,  $\f_u$ is the \sl obstruction \rm to energy-momentum conservation. The spacetime components of energy-momentum and force are obtained by choosing a local coordinate basis of vector fields $u=\pl_\a \ (\a=0,1,2,3)$.

Let us collect our expressions for $n=4$ and $p=2$ in the form
\begin{align}
\S_u(F,G)&=\frac12(F\w uG-G\w uF)\nt\\
f_u(F,G)&=uF\w dG-uG\w dF\lab{fu0}\\
\f_u(F,G)&=\frac12(F\w \5L_uG-G\w\5L_uF).\nt
\end{align}
Note that all three expressions are bilinear in $(F,G)$ and odd under the exchange $F\lra G$,
hence  invariant under the \sl electric-magnetic reciprocity \rm substitution \ci{HO3} 
\begin{align}\lab{recipr}
F\to F'=zG \hb{\ and\ }
G\to G' = -z\inv F,
\end{align}
where $z$ is any real nonzero parameter. Since $F$ and $F'$ have the dimensions of \sl action per unit charge \rm while $G$ and $G'$ have the dimensions of \sl charge, \rm $z$ has the dimensions of \sl impedance \rm (usually denoted in textbooks by $Z$). Furthermore, to keep $F'$ twist-free and $G'$ twisted, $z$ must transform as a \sl pseudoscalar \rm (or scalar of \sl odd type, \rm in the language of de Rham \ci{R84}), in particular changing sign under orientation-reversing coordinate transformations.  Reciprocity is a precursor of \sl electric-magnetic  duality \rm in the absence of a metric, as explained below. Its manifestation can be seen in the subtle structure of the extended Lorentz force, which states that the field  $F$, generated by the magnetic current $K$, exerts a force on the electric current $J$ which generates $G$, and $G$ in turn exerts a force on  $K$, as depicted in Figure 1. 

\begin{figure}[ht]
\begin{center}
\includegraphics[width=1.2 in]{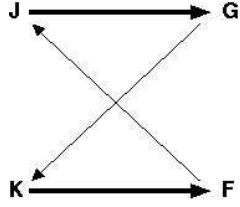}
\caption{\bf Reciprocity in action. \rm The field $F$ generated by the magentic current $K$ acts on the electric current $J$, and the field $G$ generated by $J$ in turn acts on $K$.}
\end{center}
\end{figure}

To complete the picture, recall that the Maxwell system is underdetermined and must be supplemented with constitutive relations before a solution can be contemplated. In the present context, this takes the general form \ci{P62}
\begin{align}\lab{constit}
G=G\0F,
\end{align}
which includes nonlinear or even nonlocal relations that may be inconsistent with the existence of a local metric but nevertheless describe electromagnetic phenomena in media. Requiring  \eq{constit} to be  linear, local, symmetric, and to preserve electric-magnetic reciprocity invariance generates a light-cone structure on spacetime, which determines a metric up to a conformal factor \ci[section D.6.1]{HO3} and reduces \eq{constit}  to
\begin{align*}
G=Z\inv*\!F+\a F,
\end{align*}
where $*F$ is the \sl Hodge dual \rm of $F$ \ci{F1} according to a reference metric in the conformal class, $Z$ is a conformal factor (like $z$, a pseudoscalar field with dimensions of admittance) and $\a$, called an \sl axion field, \rm is also pseudoscalar.  If we further assume that $\a=0$ and choose the metric in the class with constant $Z=Z_0$, this reduces to 
\begin{align}\lab{ML}
G=Z_0\inv *F
\end{align} 
which is the \sl Maxwell-Lorentz spacetime relation \rm describing electromagnetic propagation in a (possibly curved) vacuum spacetime if $Z_0$ is interpreted as the \sl vacuum impedance \rm 
$\sr{\m_0/\e_0}$ in SI units.  From  \eq{ML} it follows that 
\begin{align*}
2Z_0\,\f_u=F\w\5L_u {*F}-\5L_u F\w{\,*F}
=F\w {*\5L_u} F-\5L_u F\w{*F}=0,
\end{align*}
hence $\S_u$ is a `potential' for $f_u$:
\begin{align*}
f_u=d\,\S_u\,.
\end{align*}
The same argument holds for general $n$ and $p$, with some sign differences.

\section{Pre-metric complex structure}

The Hodge duality operator associated with a Lorentzian metric defines a \sl complex structure \rm on the space $\O^2$ of 2-forms, \ie
\begin{align}\lab{cs}
*: \O^2\to\O^2,\qq **F=-F.
\end{align}

Equations \eq{recipr} show that reciprocity is a precursor of Hodge duality, \ie that
\begin{align}\lab{nocs}
F'\=zG\sim *F,\qq G'\=-z\inv F\sim *G.
\end{align}
It is tempting to suppose that reciprocity actually \sl defines \rm a complex structure on 2-forms by writing
\begin{align}\lab{HO}
\cast F=zG,\qqq \cast G=-z\inv F,
\end{align}
as was done by Hehl and Obukhov in \ci{HO4}. However, this is \sl incorrect. \rm In \eq{HO}, we must know both $F$ and $G$ to compute their images under $\cast$; there is no way of defining $\cast H$ for an arbitrary 2-form $H$.
Instead, reciprocity defines a complex structure on the \sl product space \rm $\O^2\times\2\O^2$, where $\O^2$ is the space of 2-forms and $\2\O^2$ is the space of twisted 2-forms. Namely, 
\begin{align}\lab{recip2}
\cast_z(F,G)=(zG, -z\inv F) \imp \cast_z^2(F,G)=-(F, G),
\end{align}
where we have included the pseudoscalar $z$ in the notation for the operator $\cast_z$. In fact, the definition \eq{recip2} descends to the \sl tensor product \rm
\begin{align}\lab{O22}
\2\O^{2,2}=\O^2\otimes\2\O^2
\end{align}
since the latter is invariant under transformation on $\O^2\times\2\O^2$ of the type
\begin{align}\lab{k}
(F,G)\mapsto(kF, k\inv G),\qq k\ne 0.
\end{align}
The eigenvectors of $\cast_z$ are the \sl self-reciprocal \rm pairs
\begin{align}\lab{evs}
(F\9\pm_z, G\9\pm_z)=(F,G)\mp i\cast_z(F,G)&=(F\mp izG, G\pm iz\inv F),
\end{align}
with
\begin{align}\lab{evs2}
\cast_z(F\9\pm_z, G\9\pm_z)=\pm i(F\9\pm_z, G\9\pm_z).
\end{align}
Note that
\begin{align*}
F\9\pm_z=\mp iz G\9\pm_z\,,
\end{align*}
which is not surprising since the 2-dimensional \sl real \rm space spanned by the pair $(F, G)$ is equivalent to a one-dimensional \sl complex \rm space. When a Lorentzian metric is chosen and the vacuum relations \eq{ML} are satisfied, the self-reciprocal pairs with $z=Z_0$ become 
\begin{align}
(F\9\pm_z, G\9\pm_z)=(F\mp i*F, G\mp i*G)=(F\9\pm, G\9\pm),
\end{align}
where $F\9\pm$ and $G\9\pm$ are the \sl self-dual \rm forms with respect to the Hodge operator of the given metric,
\begin{align*}
*F\9\pm=\pm iF\9\pm,\qq *G\9\pm=\pm iG\9\pm.
\end{align*}

To summarize, we have two levels of complex structure. 

\bul Given \sl only \rm a differentiable 4-manifold $M$,  there is a one-paramater family of complex structures $\cast_z\ (z\ne 0)$ on $\O^2\times\2\O^2$ preserving the kinematical energy-momentum form $\S_u$ and the extended Lorentz force $f_u$. 

\bul Adding a Lorentzian metric on $M$ (so that the vacuum relation \eq{ML} is satisfied) and choosing $z=Z_0$, the reciprocity operator $\cast_z$ \sl factors \rm into the product of two Hodge operators:
\begin{align}\lab{factor}
\cast_z(F, G)=(*F, *G).
\end{align}

\subsection*{Acknowledgments} \noindent
I thank Fred Hehl and Yuri Obukhov for stimulating discussions, and Arje Nachman 
for support under AFOSR Grant \#F49620-01-1-0271.

 \end{document}